\documentclass[10pt,journal,compsoc]{IEEEtran}
\usepackage[normalem]{ulem}

\ifCLASSOPTIONcompsoc
  \usepackage[nocompress]{cite}
\else
  \usepackage{cite}
\fi

\usepackage{xcolor}
\usepackage[pdftex]{graphicx}
\usepackage{verbatim}

\usepackage[T1]{fontenc}

\ifCLASSINFOpdf
\else
\fi
\begin{document}
%
\title{IBM Functional Genomics Platform, A Cloud-Based Platform for Studying Microbial Life at Scale}
%
%
%
%

\author{
Edward~E.~Seabolt*,
Gowri~Nayar*,
Harsha~Krishnareddy,
Akshay~Agarwal,
Kristen~L.~Beck,
Ignacio~Terrizzano, 
Eser~Kandogan, 
Mark~Kunitomi,
Mary~Roth,
Vandana Mukherjee,
and~James~H.~Kaufman

\IEEEcompsocitemizethanks{
                        \IEEEcompsocthanksitem
                        * indicates co-first authors
                        \IEEEcompsocthanksitem E-mail: Gowri.Nayar@ibm.com
                        \IEEEcompsocthanksitem IBM Research, Almaden. 650 Harry Rd., San Jose, CA 95120 
                        }
\thanks{Manuscript submitted December 2019}}

%
%

\markboth{IEEE/ACM Transactions on Computational Biology and Bioinformatics ,~Vol.~16, No.~7, December~2019}
{Shell \MakeLowercase{\textit{et al.}}: Bare Demo of IEEEtran.cls for Computer Society Journals}
%



\IEEEtitleabstractindextext{%
\begin{abstract}
The rapid growth in biological sequence data is revolutionizing our understanding of genotypic diversity and challenging conventional approaches to informatics. With the increasing availability of genomic data, traditional bioinformatic tools require substantial computational time and the creation of ever-larger indices each time a researcher seeks to gain insight from the data. To address these challenges, we pre-computed important relationships between biological entities spanning the Central Dogma of Molecular Biology and captured this information in a relational database. The database can be queried across hundreds of millions of entities and returns results in a fraction of the time required by traditional methods. In this paper, we describe \textit{IBM Functional Genomics Platform} (formerly known as OMXWare), a comprehensive database relating genotype to phenotype for bacterial life. Continually updated, IBM Functional Genomics Platform today contains data derived from 200,000 curated, self-consistently assembled genomes. The database stores functional data for over 68 million genes, 52 million proteins, and 239 million domains with associated biological activity annotations from Gene Ontology, KEGG, MetaCyc, and Reactome. IBM Functional Genomics Platform maps all of the many-to-many connections between each biological entity including the originating genome, gene, protein, and protein domain. Various microbial studies, from infectious disease to environmental health, can benefit from the rich data and connections. We describe the data selection, the pipeline to create and update the IBM Functional Genomics Platform, and the developer tools (Python SDK and REST APIs) which allow researchers to efficiently study microbial life at scale.
\end{abstract}

\begin{IEEEkeywords}
Genotype, Phenotype, Relational Database, Bacterial, Microbial, Large Scale Analytics, Computational Biology, Genomics, Microbiology, Function, SDK, API
\end{IEEEkeywords}
}

\maketitle
%

\IEEEdisplaynontitleabstractindextext

%
\IEEEpeerreviewmaketitle

\IEEEraisesectionheading{\section{Introduction}\label{sec:introduction}}
\label{sec:intro}
Widespread adoption of next-generation sequencing (NGS) \cite{schuster2007} has made it inexpensive and routine to obtain sequence data. Libraries of open source tools are available to process this data for a diverse set of goals. For common microbial applications, these goals include genome assembly, alignment to references, naming organisms (based on conserved sequences), and identifying genes (annotation) \cite{Trim:2018,Flash:2018,Spades:2018,QUASH:2018,Bowtie:2018}. From assembled whole genomes, scientists can classify the genotype of an organism \cite{kent2002, wood2014}. However, genotypic information does not by itself indicate an organism's capacity for biological activity or its phenotype \cite{purcell2018,alberts2014}.  

Within proteins are sub-regions called structural and functional protein domains that perform the actual enzymatic or biochemical function(s) of the protein. Libraries of software have been developed to identify protein domains including the  InterProScan applications\cite{InterproScan:2018}. These applications also associate standardized functional codes and ontologies with specific domains to link them directly to molecular function, cellular structure, biological processes, biochemical pathways, and more \cite{ashburner2000, kanehisa2000, InterproScan:2018}. Large databases have been created to catalog domains and their functional codes, but work still needs to connect sequence to domain entities at scale through convenient application interfaces \cite{sonnhammer1997, uniprot2016}.

Using current tools, a researcher seeking to link genotype to phenotype for a particular genome must manage a large number of text files created by existing and often disparate software artifacts. These include files containing raw sequence reads, artifacts from assemblies, annotated genes, proteins, etc. The data represented is also likely to have identifiers that are unique to the source repository and do not necessarily connect data types across references. Additionally, since important genes with identical sequences may exist more than once in a genome or collection of genomes, these files contain redundant data. This traditional approach can consume vast amounts of storage and compute resources, even for relatively small bioinformatic analyses. Thus, current bioinformatic methods need to evolve in order to address the demands of processing ever-larger collections of genomic data.

In this paper, we report an application of big data techniques and relational database technologies that demonstrate a scalable approach to bioinformatics. We call the system \textit{IBM Functional Genomics Platform}. At its core, the platform is built on a database connecting genotype to phenotype for bacterial life. The data itself is derived from a growing collection of hundreds of thousands of self-consistently assembled, curated, public genomes and their associated metadata. The large magnitude of data is necessary to capture the natural sequence diversity across microbial genera. The system also provides a powerful programming and development framework to query data and to answer questions of microbial life at scale.  

Leveraging existing annotation tools \cite{Prokka:2018} \cite{InterproScan:2018}, we identify biological entities from bacterial whole genome sequence data. The entity relations and schema are based on the actual relationships between biological objects and their properties. These entities and relations are stored in a relational database to optimize storage and query time. Construction of the database itself is a compute and memory intensive process that runs on the cloud. Over 10 million compute hours went into initial construction of the database. Here we report the pipeline used for the assembly and annotation process, the database itself, indices created to optimize performance, and applications highlighting the discovery power of the IBM Functional Genomics Platform. In addition, developer tools (Python SDK, REST API) have been implemented to provide a collaboration platform that supports scientific research and the development of custom applications.

The system described is distinguishing in its breadth of data. Through our data curation and selection methods (Section \ref{sec:curation}), we ingest genomes that can be annotated with high fidelity for downstream genes, proteins, and domains. While other such repositories exist, such as Genome Taxonomy Database \cite{GTDB}, Ensembl Bacteria \cite{Ensembl}, NCBI RefGene\cite{RefSeq:2018}, MicrobesOnline \cite{microbesOnline}, and the Joint Genome Institute IMG/M \cite{JGI}, our approach results in more scaleable access to a larger corpus of genetic information (Figures \ref{graph:genomeComparison} and \ref{graph:geneComparison}). Furthermore, the ability to connect genotype to phenotype is a distinguishing factor, as we maintain connection between all relevant biological entities. This allows a user to traverse the central dogma of biology easily within one tool. Our methods provide the user with easier access to large scale multi-ohmic data and saves computational time and storage.


\section{Data Description}
\label{sec:data}
\subsection{NCBI Sequence Data}
The National Center for Biotechnology Information (NCBI) maintains a large, public domain repository of raw sequence data sets in their Sequence Read Archive (SRA) \cite{leinonen2010sequence}. This served as the source of originating genomic data for the system. To populate the initial corpus of IBM Functional Genomics data, genomic sequence data from 159,628 curated SRA data sets passing our selection criteria (described in Section \ref{sec:curation}) were combined with 6,781 complete genomes from the NCBI Reference Sequence (RefSeq) genome collection. All assemblies were based on Illumina paired-end whole genome sequence (WGS) bacterial isolate data. For our initial corpus, this selection provided 166,409 bacterial genomes representing 517 genera of bacteria. These numbers have grown to over 200,000 genomes and over 1,000 genera at the time of this publication as our \textit{NCBI monitor} (Section \ref{ncbi-monitoring}) continuously adds new genomes to the database. To ensure the structure requirements of the database (Section \ref{sec:schema}), quality control rules and curation process are further described in Methods Section \ref{sec:methods}. Genome accessions representing the initial data sets are available in the supplement. 

\subsection{NCBI Metadata}
From the NCBI BioSample database, we retrieved descriptive information about the genomic data stored within NCBI. We downloaded all metadata related to bacterial genomes within NCBI, including BioSample and BioProject, among others. For a single bacterial genome data set, there are metadata fields that describe characteristics such as isolation source, collection geography, contributing wet lab, etc. We ingested this into the database and indexed metadata from each of the NCBI reference databases, including BioProject, BioSample, GenBank, Pathogen, RefSeq, SRA, and Taxonomy. Due to the undefined naming conventions (e.g. over 1500 attributes from only BioSample), we did not impose a structure to the metadata but rather ingested and provided access to all fields from the seven sources. 

The NCBI metadata is important to the IBM Functional Genomics Platform as it allows for experimental classification of sequence data. The metadata artifacts are related to the biological entities, and they can be essential sources of ground truth data in downstream analysis. For example, indexing BioSample antimicrobial assay data allows us to layer on the antimicrobial resistance phenotype with biochemical confirmation to the related genome data. Alternatively, one could use the metadata to filter a set of genomes from a given isolation source and then observe enriched gene or protein signatures. It is crucial to ingest metadata along with the sequence data to support experimentally-informed analysis.

\subsection{Breadth and Depth of Biological Data}
Next-generation sequencing has been applied to the characterization of microbial life for over a decade; yet, we have only cataloged a fraction of the total microbial diversity present on Earth. Projects such as the Earth Microbiome Project \cite{earthmicrobiomeproject}, Human Microbiome Project \cite{humanmicrobiomeproject}, 100K Pathogen Project \cite{100kpathogenproject}, and others aim to expand available sequence data describing microorganisms. These efforts are essential to our understanding of the microscopic world as microbial genotypes are dynamic. They evolve function under environmental stress e.g. when exposed to heat, pH extremes, antibiotics, etc. and will readily exchange DNA with other microorganisms in their environment. This results in an ever-expanding universe of genotypic sequence. Nevertheless, analysis in this field depends on references built from these sources; specifically, large quantities of microbial reference sequences are required to comprehensively capture the true underlying diversity\cite{insular2019Book}.

For analysis of whole genome or metagenomic data, many researchers use NCBI RefSeq microbial genomes as a standard reference. NCBI RefSeq contains 14,800 bacterial genomes (assembly level marked as "complete") as of October 2019. NCBI RefSeq Complete genomes are often considered a gold standard, and so are included in the database. However, RefSeq alone does not capture the full microbial biodiversity; and without expansion, limited references can lead to incorrect microbial analysis\cite{ice2019ASM,insular2019Book}. 

\begin{figure}[!ht]
\begin{center}
\includegraphics[trim=0cm 0cm 0cm 0cm, width=1\columnwidth]{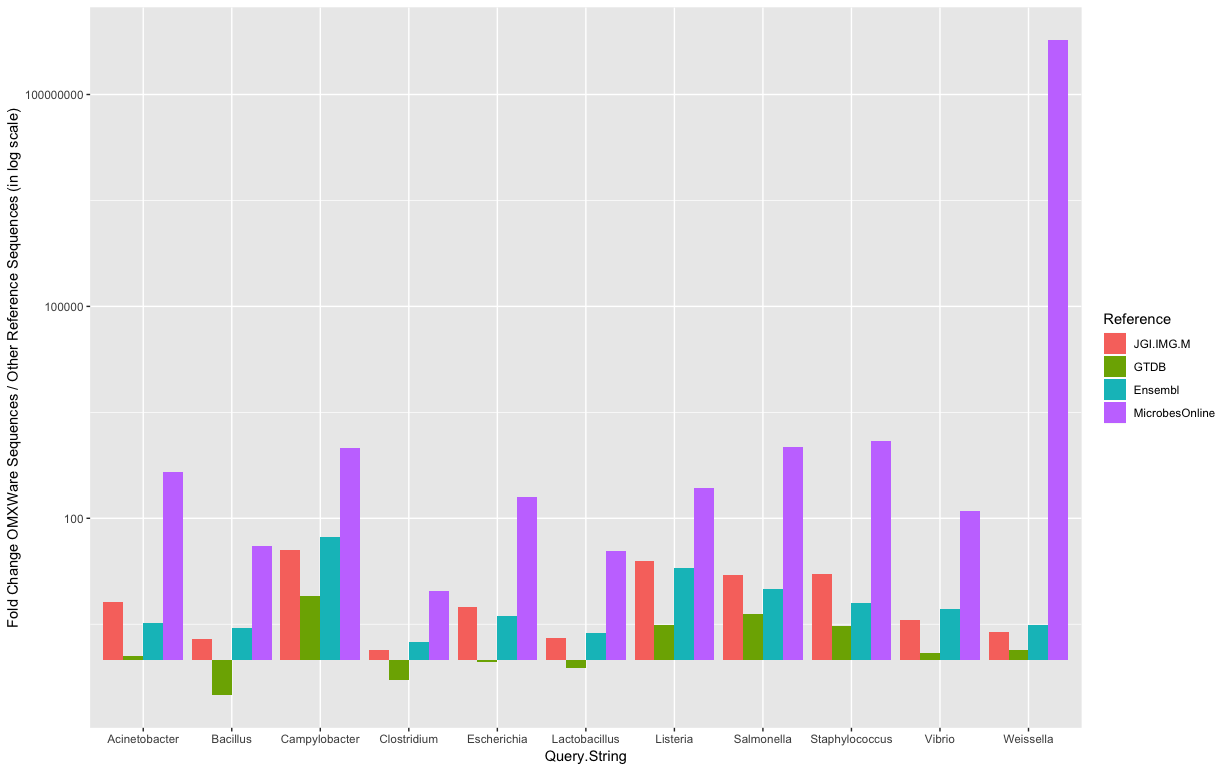}
\end{center}
\caption{Genome data availability in the described platform relative to other public references. Fold change is calculated for the number of unique genomes for a given genus within IBM Functional Genomics Platform vs. the named references and displayed in log scale.}
\label{graph:genomeComparison} 
\end{figure}

Several efforts have emerged to produce more comprehensive repositories of microbial reference data including the Genome Taxonomy Database (GTDB)\cite{GTDB}, Ensembl Bacteria \cite{Ensembl}, NCBI RefGene\cite{RefSeq:2018}, MicrobesOnline \cite{microbesOnline} (recommended for microbiologists by Zhulin et al. \cite{databaseForMicrobiologist}), and the Joint Genome Institute IMG/M \cite{JGI}. 

\begin{figure}[!ht]
\begin{center}
\includegraphics[trim=0cm 0cm 0cm 0cm, width=1\columnwidth]{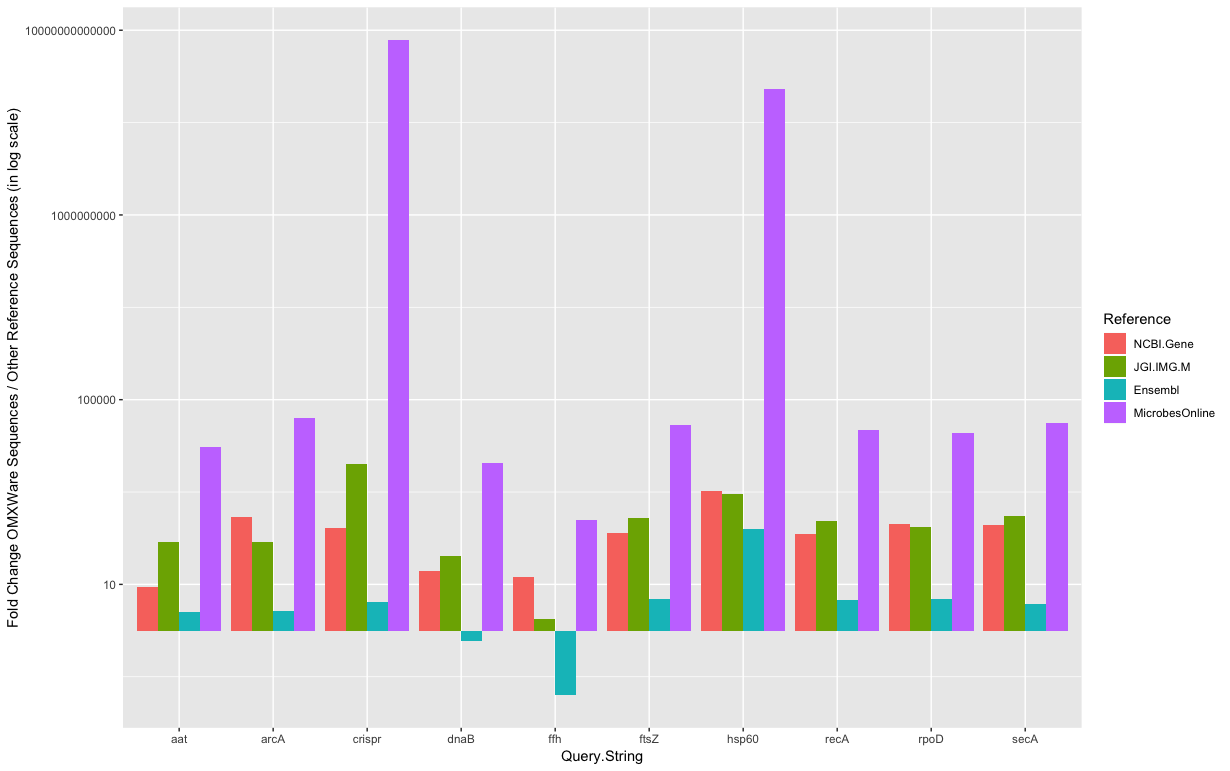}
\end{center}
\caption{Gene data availability in the proposed platform relative to other public references. Fold change is calculated for the number of unique gene sequences for a given gene name substring within the platform vs. the named references and displayed in log scale.}
\label{graph:geneComparison} 
\end{figure}

To compare IBM Functional Genomics Platform with these references, we selected a core set of genera and genes and subsequently retrieved the corresponding entries from each reference source. The count of genome assemblies across these rare or common genera and the total sequences for a gene name substring are reported here (Figure \ref{graph:genomeComparison} and \ref{graph:geneComparison}). As described in Section \ref{sec:bio_data_scaling}, we store each genome, gene sequence, and protein sequence by unique identifiers (accessions for genomes and md5 hash of the sequence for genes and proteins), and each unique entity appears exactly once within the database. Therefore, the counts reported here show the number of unique examples within each genus and gene name within the database.  Figures \ref{graph:genomeComparison} and \ref{graph:geneComparison} show the fold change of sequences in the database relative to each reference for genome and gene data. The median fold enrichment is 5x and 174x for genome and gene data, respectively, in the described platform compared to other references (Figures \ref{graph:genomeComparison}-\ref{graph:geneComparison}). Similar comparisons can be done based on the number of annotated domains present in other repositories. For example, IBM Functional Genomics Platform has 35x more functional domains than M5nr\cite{m5nr}--- 233,418,698 domains compared to 6,626,200 in the M5nr database. Our system uses the same five largest reference sets as M5nr, specifically GenBank, KEGG, GO, UniProt, and RefSeq, to annotate the functional domains. However, the main distinguishing feature, is the maintenance of relationships from the functional domains to the parent protein sequence, gene sequence, and ultimately originating genome. M5nr does not support such connections and only tracks the originating organism for the functional domain and associated metadata. Furthermore, the M5nr database is not accessible directly from the cloud, and thus a user must first download the dataset before data can be retrieved and analyzed. We describe in Section \ref{sec:userInterface} the programmatic and web-based methods we have implemented that leverage the power of a cloud-based system to allow a user to access data with limited installations required.

\section{Methods}
\label{sec:methods}
In the following section, we describe the process of selecting and curating the genomic data from public sources, to annotation of genes, proteins, and functional domains. Our selection and curation methods described below are implemented in order to ensure that downstream genes and proteins can be annotated, and thus we select data with the purpose of connecting the genome to phenotype. Each entity resulting from the assembly and annotation methods are stored within the database, referenced by a unique identifier, and their connections are preserved. Thus, this methodology allows the system to connect the genome sequence to the phenotype of the organism.

\subsection{Data Selection}
\label{sec:selection}
The quality of WGS and accuracy of metadata maintained by NCBI varies dramatically. For instance, NCBI includes some data sets derived from contaminated samples (not pure bacterial isolates).  Other microbial data sets have been assigned incorrect taxonomic identifiers, which compromises identification accuracy in downstream analyses. Additionally, some sequenced isolates do not provide sufficient depth or quality of sequencing to adequately represent the genotype. To address these deficiencies, IBM Functional Genomics Platform systematically curates NCBI’s microbial WGS data. 

IBM Functional Genomics Platform genomic data is populated from four main sources in NCBI: GenBank, Pathogen, RefSeq, and Sequence Read Archive (SRA). The latter of which requires assembly of raw sequence data and additional curation/processing. To identify whether an assembled genome from GenBank, Pathogen, or RefSeq should be added to the database, the following conditions must be met:
\begin{enumerate}
\item Using the taxonomic tree, we identify if the genome is of bacterial lineage.
\item We identify if the genome has an assembly level of "Complete Genome." 
\end{enumerate}

For SRA or unassembled Pathogen genomes, the process is more complicated as the raw reads must go through an assembly process first. The conditions for adding or updating this data are as follows:
\begin{enumerate}
\item Must be bacterial as defined by the data set's taxonomic lineage.
\item Library strategy must be Whole Genome Sequenced (WGS).
\item Library source must be Genomic.
\item Library descriptor must indicate pair-ended reads.
\item Library platform must be Illumina (selecting short read sequencing only).
\end{enumerate}

The requirement of either "Complete Genomes" (for GenBank and RefSeq accessions) or "Illumina pair-ended read" (for paired-end reads) ensures that SPADES delivers an assembly of the highest quality possible as measured by N50 and number of contigs, described in Section \ref{sec:curation}. Since we store the annotated genes, proteins, and domains as associated database entities for each genome, a poor assembly is most likely to impact the database if that assembly fails to detect genes. While our ingestion pipeline does not prevent using incomplete genomes, in order to maintain the structure of the database, described in Sections \ref{sec:entities} and \ref{sec:relations}, we use the above requirements to select genomes that can be annotated for gene and protein sequences. 

Data sets matching these criteria were converted to FASTQ format using the NCBI SRA Toolkit Technology \cite{sherry2012ncbi}. To populate IBM Functional Genomics Platform with the initial corpus of data, over 360,000 raw WGS data sets met these initial criteria and were filtered further after assembly as described in Section \ref{sec:curation}.

For any of these data sources if the above conditions are met, the genome will progress through the genome assembly (if needed) and annotation steps described in Sections \ref{genome_assembly} and \ref{geneProteinAnnotation}.

\subsubsection{NCBI Monitoring} 
\label{ncbi-monitoring}
IBM Functional Genomics Platform is not just a static database, but increases as the available bacterial information grows. NCBI reports an average annual growth rate of $36.9 \%$ of bacterial genetic data\cite{ncbiGrowth}. Regulatory agencies such as the US Food and Drug Administration and Center for Disease Control submit sequenced isolates relating to infectious disease outbreaks from across the country daily. Additionally, it is typically required to submit raw sequence data for any biological publication. Thus, we devised a new monitoring service to continually update the database with the latest genome assemblies and sequence data from NCBI. The process allows the database to contain the most current data and optimizes for high-quality data that passes our previously stated curation thresholds. 

The platform monitors seven key databases from NCBI: BioProject, BioSample, GenBank, Pathogen, RefSeq, Sequence Read Archive (SRA), and Taxonomy. Each genome is associated with its corresponding NCBI metadata record from these sources to provide additional experimental information and to aid our users in discovering relevant data. BioProject and BioSample are used to supplement and enrich biological entity search capabilities. GenBank, Pathogen, RefSeq, and SRA are used to continually update the bacterial genomic sequence database and yield their downstream genes, proteins, and functional domains. Microbial nomenclature and taxonomy can change as new genotypic and phenotypic evidence is discovered; therefore, we continually update to the latest NCBI Taxonomic Tree to ensure our genomic data is associated with the most current identifiers known to date.

If the data selection requirements are satisfied, the monitor will schedule the assembly (if required) and annotation pipeline to run on the genome. This includes retrieving or generating the assembled genome data and annotating the genes, proteins, domains, pathways, gene ontology, etc. The annotation pipeline is described in Section \ref{geneProteinAnnotation} and \ref{domainAnnotation}.

These monitoring processes are crucial in maintaining the relevance of data. It is important to update the database with the most current information and metadata, as biological knowledge grows with more sequencing data. Furthermore, with the expanding amount of sequence data, an updated relational database becomes more useful, as the methods explained above allow for optimized genome, gene, protein, and domain analysis that would be otherwise intractable. Sections \ref{sec:bio_data_scaling} and \ref{sec:omxware_apps} further elaborate on these capabilities.

\begin{figure}[!ht]
\begin{center}
\includegraphics[trim=0cm 0cm 0cm 0cm, width=1\columnwidth]{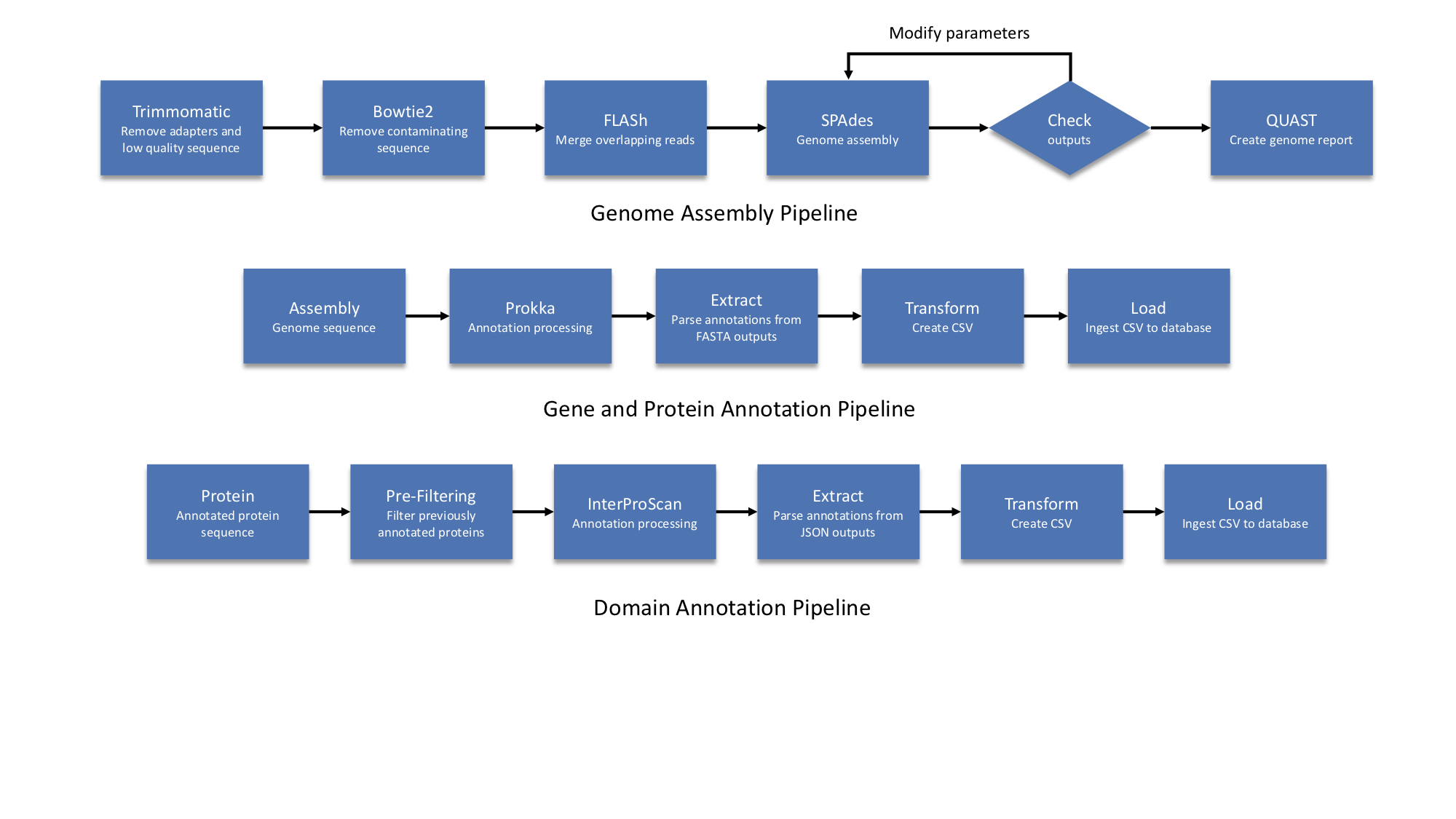}
\end{center}
\caption{Assembly, annotation, and curation pipelines}
\label{Pipeline} 
\end{figure} 

\subsection{Genome Assembly} 
\label{genome_assembly}

After the criteria in our selection process is applied to the data set, raw sequence data from SRA or Pathogen is then processed for assembly and annotation. Figure \ref{Pipeline} represents the individual stages executed in the assembly pipeline. To ensure high-quality data for the assembly, several pre-assembly steps were performed including removal of low-quality reads and contamination from sequencing adapters with Trimmomatic \cite{Trim:2018} as well as alignment and removal of PhiX (NCBI Reference Sequence: NC\textunderscore001422.1)  with Bowtie2 \cite{Bowtie:2018}. PhiX is an internal sequencing control and common contaminant in this type of data. In order to reduce the number of unassembled contigs and increase N50 value, we used FLASh \cite{Flash:2018} to merge paired-end reads and to improve the overall quality of the resulting assembly. 

\begin{figure}[!ht]
\begin{center}
\includegraphics[trim=0cm 0cm 0cm 0cm, width=1\columnwidth]{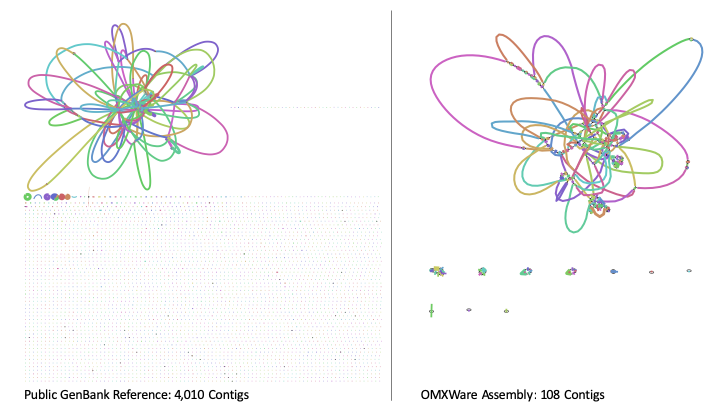}
\end{center}
\caption{Exemplar comparison of an \textit{Acinetobacter} genome assembly in pubic domain and IBM Functional Genomics Platform. This figure shows two bandage plots comparing assemblies from the same raw sequencing data: public domain GenBank assembly GCA\textunderscore001696615.1 (\textit{left}) versus IBM Functional Genomics Platform assembly of SRA data (\textit{right}, assembly of raw data from SRR3938306).  The connected lines within the bandage plots indicate regions within an assembly that map to contiguous regions (contigs). Very short contigs appear as symbols (or points) below the plot. A more complete assembly has fewer contigs and is more representative of biological truth.}
\label{bandage-plot} 
\end{figure}

Once pre-assembly steps were complete, the merged reads were assembled in an iterative assembly/quality evaluation process with SPAdes and QUAST to optimize for the most complete genome assemblies\cite{Spades:2018, QUASH:2018}. This dynamic processing allowed each isolate to use assembly parameters optimized for the data type and to yield the highest quality assembly. By adding this novel step to our pipeline, we were able to improve assembly quality significantly. For instance when considering an exemplar \textit{Acinetobacter} data set with public genome assembly (Genbank ID: GCA\textunderscore001696615.1) and available raw sequence reads (SRA ID: SRR3938306), we're able to compare the continuity of assemblies resulting from two pipelines using the same starting sequencing data (Figure \ref{bandage-plot}). The public assembly consists of 4,010 contigs whereas our assembly of the same raw sequencing data yielded a more continuous assembly of 108 contigs (data retrieved June 2017). Furthermore, our genome assembly yields a higher N50 of 116,650 bp compared to the public Genbank assembly N50 of 74,381. This indicates a 1.57-fold improvement in contig length and continuity which for our assembly is supported with a median coverage value of 24X.  Additionally, the L50 of the Genbank assembly is 19 contigs compared to only 11 from our assembly pipeline. Increased assembly continuity more accurately represents true bacterial genome structure and provides better starting data for gene, protein, and functional annotation.  After assembly, the genomes resulting from SRA data were then further processed for taxonomic naming accuracy as described in Section \ref{sec:curation}.

\subsection{Curation of Assembled Genomes}
\label{sec:curation}

Most bacterial genomes contain one circular chromosome and a small number of plasmids. Genome assemblies based on fewer, longer contiguously \textit{de novo} assembled regions more accurately represent the true underlying genome structure compared to assemblies comprised of larger numbers of short contigs. More contiguous assemblies also support higher quality annotations of genes, proteins, and functional domains. Therefore, we discard contigs less than 500 bp in length. Additionally, to optimize for the best genome coverage, the assembly process prioritizes selecting the assembly with the highest N50 value using an an iterative and dynamically generated k-mer set size. This malleable process is better able to adapt to natural differences in genome size and GC-content (or guanine-cytosine content) across various microorganisms and achieve overall better quality assemblies. From the original corpus of SRA datasets, only approximately 48\% bacterial genomes met the aforementioned curation thresholds.

As previously mentioned isolate contamination, called chimerism, or mislabeling of genomes can occur; therefore, we employed a strategy to ensure pure single bacterial isolate origin and a valid genus level classification. A Kraken reference was built from 6,781 RefSeq Complete genomes using the NCBI taxonomy (retrieved April 2017) \cite{wood2014, RefSeq:2018} and subsequently used to screen the newly assembled genomes in the original corpus.  With this reference, k-mers from each assembled genome were classified on a contig-by-contig basis to identify the lowest common ancestor (LCA) in the taxonomy tree. The resulting k-mer hits per node were rolled up to the genus level. This analysis provided a measure of both the k-mers matched to a genus i.e. new k-mers not in the reference ($N_{new}$ k-mers) and k-mers classified to multiple genera. The latter would lead to information loss at the leaf nodes if the new genome was added to the database.  For the majority of genomes, all the k-mers matched exclusively to one genus. However, some genomes contained k-mers ($N_{otherGenus}$) that matched a single genus different from the labeled genus or matched two or more genera. The first case is evidence of NCBI-contributor mislabeling. Adding such a genome would degrade the performance and accuracy of the reference database. The second case indicates possible isolate contamination or chimerism. 

For those genomes causing significant roll up of k-mers to a higher LCA i.e. a larger amount of $N_{otherGenus}$ k-mers, we evaluated the ratio of information gained to information lost by the addition of a genome to the reference. All genomes with a ratio of larger than 20:1 gained to lost k-mers were added to the original RefSeq Complete Kraken reference, and all other genomes were set aside as indeterminate genomes. The indeterminate genomes were then reclassified on the now larger Kraken reference on a contig-by-contig basis, using the same ratio threshold to determine if the genome should be included. From this analysis, we found that 159,628 genomes were well-classified representatives of their designated genus. In combination with the original 6,781 RefSeq Complete genomes, this yielded a high-quality collection of 166,409 bacterial genomes which served as the initial seed data of IBM Functional Genomics Platform. By curating genomes in this manner, 13,044 genomes were rejected. This may have resulted in a loss of some unusual biological samples. It is possible to annotate these rejects and in the future, relate such annotated entities to accessions marked as unusual or suspect. For the initial release, we chose to exclude these unusual data sets. With these curation steps, biological entities in IBM Functional Genomics Platform are related to classified genomes with a vetted label at the genus-level taxonomic rank. 

\subsection{Gene and Protein Annotation}
\label{geneProteinAnnotation}
As shown in the second pipeline of Figure \ref{Pipeline}, after genome assembly, genes and proteins were annotated. First, genes and proteins were discovered from the assembly using Prokka\cite{Prokka:2018}. Next, the generated ".fna" and ".faa" files were parsed, resulting in the collation of the genome, gene, and protein data entities into CSV files. Finally, the CSV files were loaded into the appropriate tables within the database. The database schema is described in Section \ref{sec:schema}.

\subsection{Domain Annotation}
\label{domainAnnotation}
\begin{figure}[!ht]
\begin{center}
\includegraphics[trim=0cm 0cm 0cm 0cm, width=1\columnwidth]{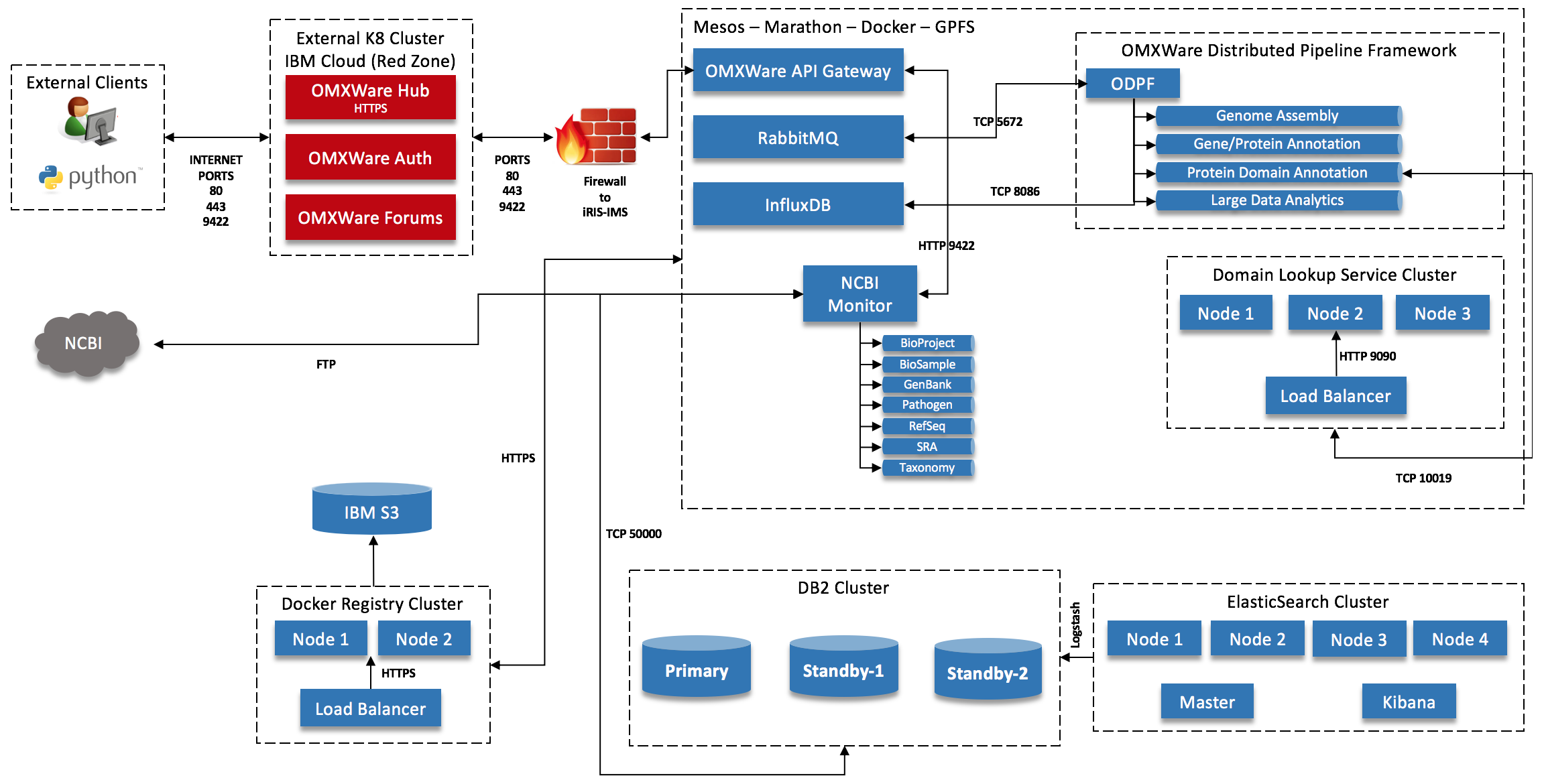}
\end{center}
\caption{OMXWare Distributed Pipeline Framework Architecture}
\label{ODPF} 
\end{figure}

In order to describe the phenotypic capability of a genome, we consider the domains of each protein contained within that genome. Protein domains are sub-strings of the protein which determine the enzymatic activity and thus deliver biological function. After gene and protein annotation, protein domains were identified using InterProScan \cite{InterproScan:2018}. The third pipeline in Figure \ref{Pipeline} illustrates this phase. First for a new genome, unique sequences for annotated proteins were scanned against the existing database to determine if they were previously identified. The output of this scan produced a reduced set of new protein sequences, which greatly decreased the amount of time required to analyze the protein sequences by InterProScan. Next, InterProScan was run on the reduced protein sequence set. All 16 available analyses provided by InterProScan were run over all input sequences and the results were output in JSON format. To reduce the amount of time spent in this analysis step, we distributed each of the 16 algorithms to individual processes via GNU parallel within the executing stage. We also leveraged InterProScan's ability to utilize a local network lookup service, which we placed in a cluster and load balanced as shown in Figure \ref{ODPF}. Next, for each of the 16 resulting JSON documents produced by InterProScan, we parsed the annotated domain information into a set of CSV files. As in the gene and protein annotation step, these CSV files are loaded into the appropriate tables in DB2. By using these steps, we were able to efficiently identify the domains within the protein sequences to connect the genotypic information to phenotype or function.

\section {Architecture}
\label{sec:achitecture}
Due to the large scale of sequence data produced by this pipeline, we implemented a cloud-based architecture to effectively orchestrate the complex use of the bioinformatic tools described above across multiple servers. We created the OMXWare Distributed Pipeline Framework (ODPF) to orchestrate the execution of the open-source components for genome assembly and annotation of the bacterial genomes retrieved with the NCBI monitor. As we annotate the various entities, genes, proteins, and functional domains, the size of the data only increases, so we use the storage advantages of a relational database to maintain the entities and relationships. Using the following schema design, we are able to store the initial and produced sequence data and metadata, saving the user compute time, as well as allow traversal between the various entities. 

\subsection{System Details}
\label{sec: systemDetails}
We leveraged the IBM Cloud to provision and deploy essential bioinformatic tools, described in Section \ref{sec:methods}. These bioinformatic analyses require large compute time, so to optimize the runtime, we used a combination of bare metal and virtual machines totaling over 1468 CPUs, 6TB RAM, and 160TB of hard drive space. With such a large number of machines running concurrently, it was necessary to implement methods for cluster management and orchestration. Therefore, we utilized Apache Mesos for cluster resource management and scheduling \cite{kakadia2015apache} and Marathon for Docker container orchestration and health checking \cite{hoque2017towards}. In order to efficiently monitor the execution and performance of the system, we used RabbitMQ as a broker \cite{dossot2014rabbitmq} to implement a message-oriented methodology for executing pipelines. This also emits system events from all pipelines, captured in the time series database InfluxDB \cite{naqvi2017time,date1984guide}. To automate management and execution of the system, we created the ODPF, to coordinate incoming messages from the broker, execute individual stages of a pipeline, record events as each stage progresses, and route messages to additional queues when requested. This was a crucial component to enable the computationally intensive task of assembling and annotating such a comprehensive set of genomes.

\subsection{Pipeline Execution Mechanics} 

A pipeline job can be started by creating a JSON document with the desired stages to execute and submitting it to the target queue. As a job is processed, the ODPF will emit events for the start, completion, and failure of stages. An event contains several fields including the context of the stage executed, environment settings from Mesos and Marathon, the host at which the process is executing, and many others. If an error is encountered during processing, ODPF will create a failure queue, denoting the name of the current queue that has failed. The failed portion of the job is sent to this queue for later inspection. Additionally, a failure event is recorded in InfluxDB which contains all the necessary information for an administrator to locate where in the cluster the failure occurred and why.

An instance of ODPF is encapsulated as a Docker container and deployed into a cluster through Marathon's management console or API. Only a minimal initial configuration is required for this deployment including access to the Docker socket file, the name of the InfluxDB database to store events, and message broker host and input queue name. The Docker socket file is required as ODPF coordinates the execution of sibling containers based on the contents of the job received from the message broker. By allowing the job descriptor to describe queue routing behavior, we can fine-tune how a job is processed. For example, some bioinformatic processes tend to take longer to reach completion than others. Letting stages or invocations define their target process queue provides the flexibility for a particular queue to be serviced by ODPF instances allocated with additional cluster resources. This allows jobs to be highly distributed without stalling the pipeline due to a slow stage.

\subsection{Biological Data Scaling}
\label{sec:bio_data_scaling}

IBM Functional Genomics Platform utilizes a relational database to efficiently store and retrieve the interconnected biological entities and sequence data. The relational database allows us to index the data, perform quick queries, and creates high read/write access. This allows for fast operations and analysis of the annotated data on a large scale. 

\begin{figure}[!ht]
\begin{center}
\includegraphics[trim=0cm 0cm 0cm 0cm, width=1\columnwidth]{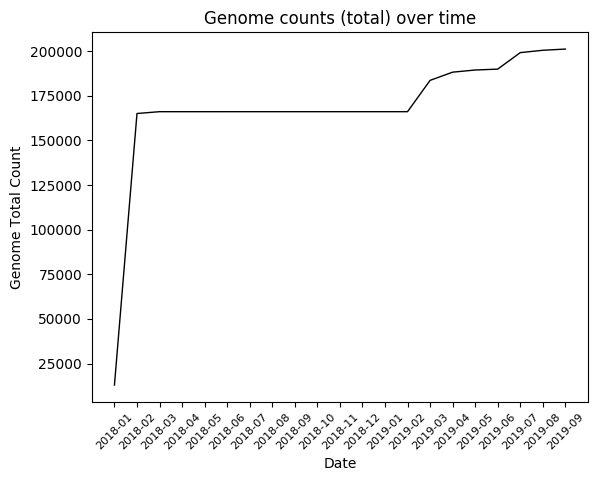}
\end{center}
\caption{Rate of genome ingestion into the database over time}
\label{genome_count} 
\end{figure} 

By using a relational database, we were able to optimize data storage, as each unique entity is stored once with no duplicates. For genomes, we use the genome accession, provided by NCBI, as the unique identifier to prevent storing duplicate genomes. If NCBI makes an update to a particular genome, we update that accession's recorded data rather than create a new instance of the same accession. For genes and proteins, we use an Md5 hash of the sequence as the unique identifier within the database in order to prevent storing duplicates. If two genomes contain the same gene sequence, instead of storing that gene sequence twice, we create one entry for that gene and maintain two unique relationships, one from each originating genome (same method is used for proteins). In this manner, we store each entity exactly once within the database, but may maintain multiple connections to a particular entity. These entities and relationships are further described in Sections \ref{sec:entities} and \ref{sec:relations}.

\begin{figure}[!ht]
\begin{center}
\includegraphics[trim=0cm 0cm 0cm 0cm, width=1\columnwidth]{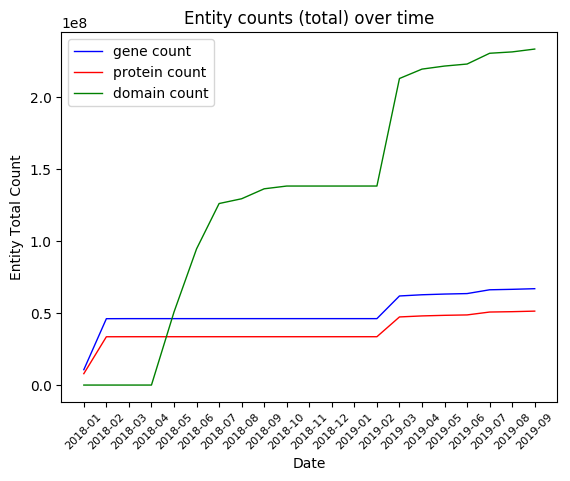}
\end{center}
\caption{Number of biological entities in database over time. This figure shows the rate of discovery of new genes (blue), proteins (red), and domain (green) sequences as they are ingested into the database.}
\label{entity_count} 
\end{figure}

As shown in Figure \ref{genome_count}, after the initial genome corpus ingestion, there is an only incremental growth rate from the NCBI Monitor (Section \ref{ncbi-monitoring}). For gene and protein entities, after initial ingestion, the growth rate was sublinear (Figure \ref{entity_count}), and we have previously shown that the rate of discovery of novel genes and proteins scales approximately to the square root of the number of genomes processed \cite{kaufman2019insular}. Upon introducing domain annotation to our pipeline, we observed a similar trend beginning with a high discovery rate that reaches a plateau. Thus, we expect a slower rate of growth across biological entities as new data is continually introduced. The long plateaus of no growth in Figures \ref{genome_count} and \ref{entity_count} reflect periods in which the ingestion of new data into the system was paused for testing.

\begin{figure}[!ht]
\begin{center}
\includegraphics[trim=0cm 0cm 0cm 0cm, width=1\columnwidth]{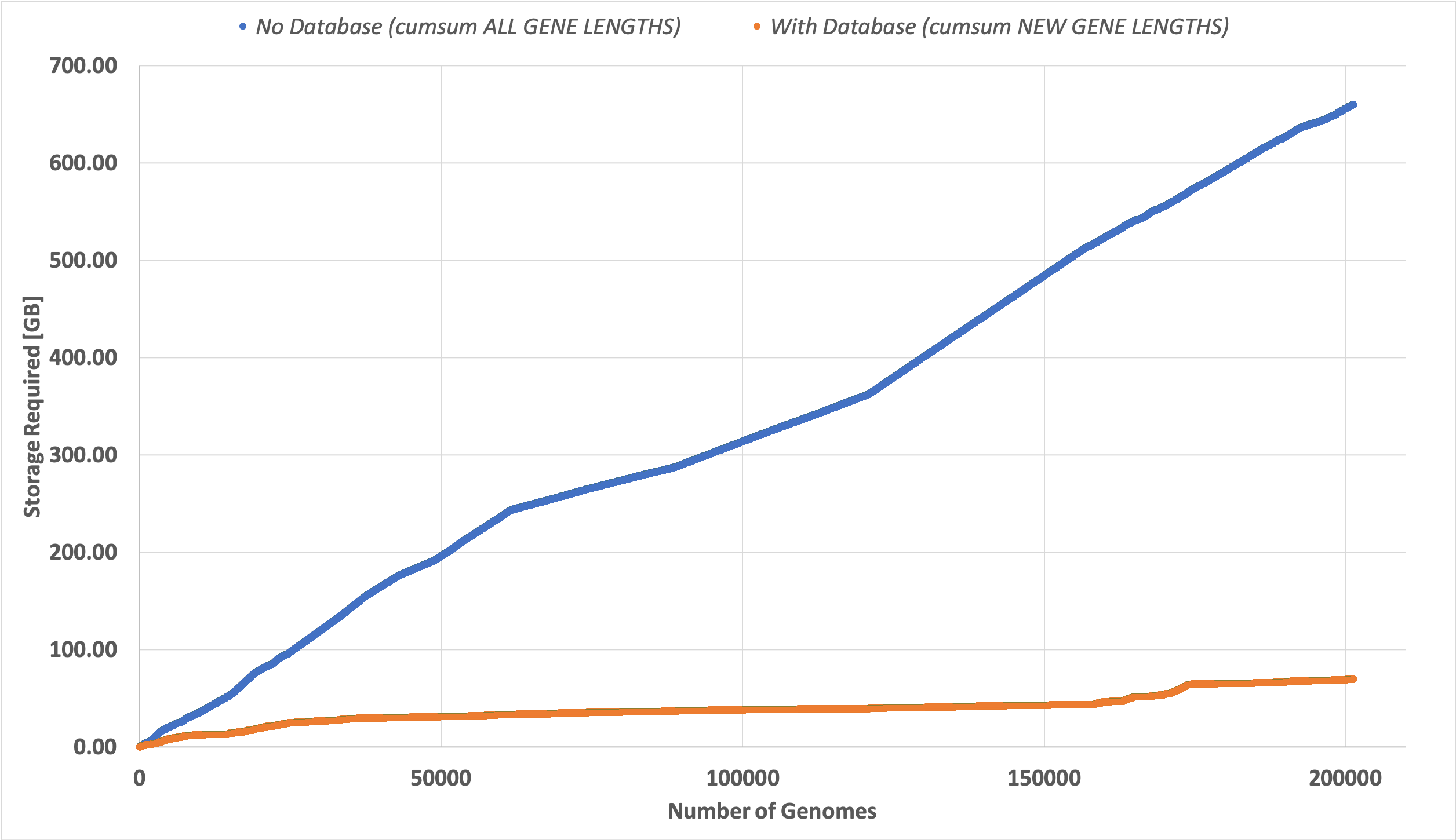}
\end{center}
\caption{Biological data scaling advantages with IBM Functional Genomics Platform. The size of sequences in the gene table grows in GB (or nt $\times 10^9$) as new genomes are added to the database (orange curve). By comparison, far greater and more rapidly growing storage would be required in a system that simply stored all annotated sequences in text files e.g. FASTA files without headers (blue curve).}
\label{db_scaling} 
\end{figure}

We note that efficiency gained from a relational data store derives from the natural frequency distribution representing the rate of discovery of gene sequences across genomes. This is evident in Figure \ref{db_scaling} which, as a function of the number of genomes processed, compares the size of sequences in the gene table to the storage requirements if the data had simply been in flat files. By leveraging a relational database, the gene table grows in single GB (or nt $\times 10^9$) as new genomes are added to the database in contrast to the far higher storage requirements (700GB) for a system that simply stored all the annotated sequences in text files e.g. FASTA files not considering sequence headers. The benefit of using a properly designed relational database further increases since unique sequences are stored only once. Specifically, Figure \ref{db_scaling} demonstrates the ~10x efficiency gain when storing ~200,000 whole genomes. The same storage efficiency gain extends to all biological entities.

\subsection{Database Schema}
\label{sec:schema}
\subsubsection{Entities}
\label{sec:entities}
The biological entities within IBM Functional Genomics Platform are the central objects stored in the database. These include genomes, genes, proteins, and domains. IBM Functional Genomics Platform has a total of 5 entities and 36 tables to capture sequence data,  metadata, and mapping data. As described in Section \ref{sec:bio_data_scaling}, we use a unique identifier for each entity, the accession for a genome and the md5 hash of the sequence for genes, proteins, and domains. Storing the entities in this manner allows for efficient storage and maintenance of all relationships from the genome to the domains with associated metadata.
\begin{enumerate}
    \item \textbf{Genome}: We store the genome accessions from NCBI and use this as a unique ID for the genome and associated assay information. We also store the validated genus and full NCBI taxonomic lineage. The genome entities are connected to the genes and proteins found within the particular genome.
    \item \textbf{Gene}: For each gene found within a genome, we store the annotated gene sequence and maintain this relationship between gene sequence and the genomes that contain it. Furthermore, we store the name of the gene assigned by Prokka and associated short names to allow users to query for genes in a traditional manner. We store the relative position of each gene on each source genome contig as reported by Prokka. The gene entities are connected to their downstream proteins. 
    \item \textbf{Protein}: Similar to genes, for each protein, we store the annotated protein sequence and maintain the relationship to the originating genome and gene sequences. We store both the full name and short name for each protein sequence and the relative position of the protein sequence on each source genome contig, as given by Prokka. The protein entities are connected to their downstream domains.
    \item \textbf{Domain}: For each domain found within a protein, we store the name and description and a connection to the originating protein. Since domains are the distinct functional or structural units on a protein, the ability to easily query this information is crucial to connect genotype to phenotype. Thus, we represent domains using IPR codes and model these IPR codes as sub-entities within the database to improve performance.
    \item \textbf{IPR}: IPR codes are assigned to protein domains by Interpro and are a representation of domains and protein families. For each IPR code, we store its associated description, name, and type. The IPR codes are linked to their corresponding domain, and this relationship can be used to map back to the original protein, gene, and genome.
     \item \textbf{Pathway}: Pathways represent molecular interactions, reactions, and relation networks for biological functions.  The database contains pathway information from three databases: KEGG, MetaCyc, Reactome. We store the pathway code, the source database, and its description. The pathway codes are related to both domains and IPR codes to allow for a fast association between the domain entities and pathways. Pathways are commonly represented using gene ontology (GO) terms.
     \item \textbf{GO}: The GO terms are representations of the pathway information described above. The database contains fields describing the terms, their name, their description, their type, and contain a direct connection to the associated IPR codes. This reduces the number of joins necessary to traverse back to the originating sequence data.
\end{enumerate}

\subsubsection{Entity Relations}
\label{sec:relations}
Once the entities and related data have been modeled and stored, it is important to connect these entities in the relational schema. Using our annotation process (Section \ref{sec:methods}), we find the genes contained within the genome. Each gene is stored as a unique entity and we maintain the relationship between the genome and its genes through a mapping table, mapping between the accession and the hashed gene sequence. Proteins are stored in a similar fashion, with mapping tables maintaining the relationship between the genome accession and the hashed protein sequence, and the hashed gene sequence and hashed protein sequence. Furthermore, in order to persist the relationship between the genes and proteins, we create a gene-protein mapping table that utilizes the locus information, provided by Prokka, in order to connect the genes and proteins. This locus id is surfaced as \textit{gene protein mapping index} within the database. It can be used to both map the genes to the corresponding proteins, as well as finding the nearest neighbors, as described in Section \ref{sec:exampleApp}.

For each protein sequence, we found the associated functional domains and store each unique functional domain found. We maintain the relationship between the hashed protein sequence to the functional domains that it contains. Therefore, we maintain both each unique entity found as well as the relationship from genome accession, to gene, to protein, to functional domain. Thus, in fact, using the entity tables and the mapping tables, we can map from genes to their function, serving to connect genotype to phenotype.

\begin{figure}[!ht]
\begin{center}
\includegraphics[trim=0cm 0cm 0cm 0cm, width=1\columnwidth]{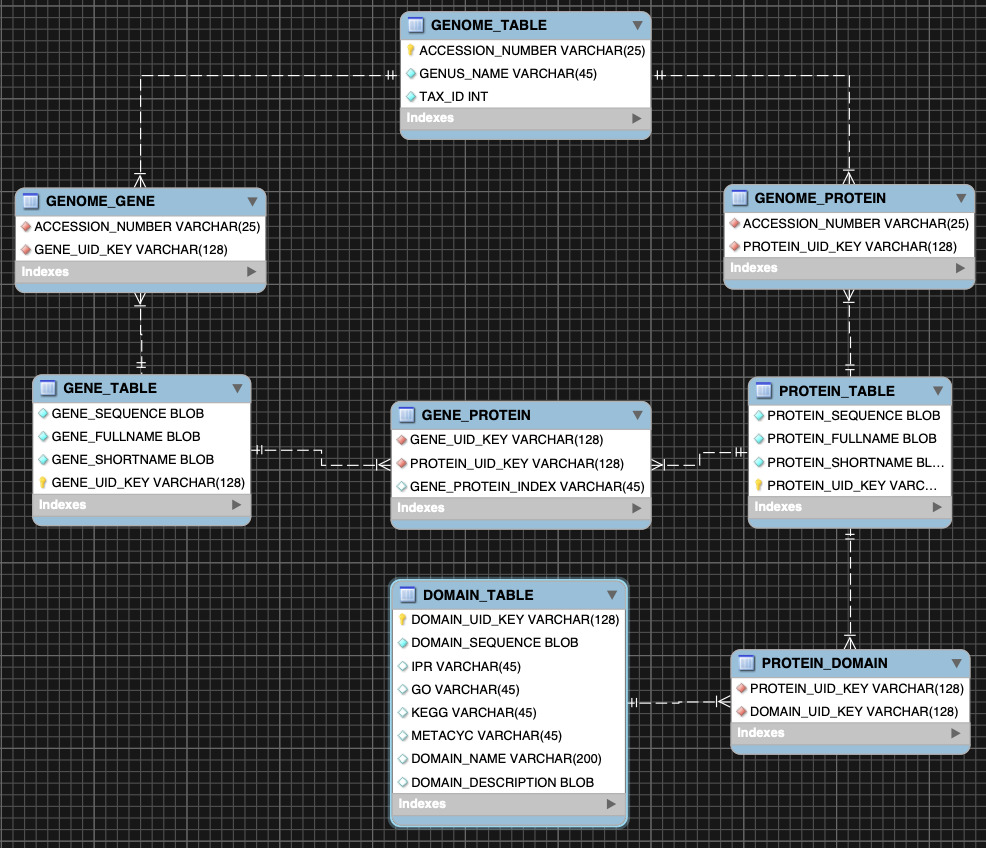}
\end{center}
\caption{Simplified Entity Relationship (ER) Diagram. A simplified ER diagram is shown representing the most important entity relations in IBM Functional Genomics Platform. For performance reasons some of the tables shown above are split into \textit{primary object} and \textit{object details} tables.}
\label{ERD} 
\end{figure}

The structure of these relationships is described in Figure \ref{ERD}, which details the data stored for each main entity and the associated relationships. We can see that this relational model is crucial in reflecting the biological relationship between genotype and phenotype. In order to allow for genomic analysis at scale, it is important to maintain these connections for all annotated sequences, rather than repeatedly perform the annotation process as needed. Such relations can allow the user to draw generalizations from the data, such as identifying genera that commonly contain a particular gene. By maintaining relationships between this pre-computed data, we are able to facilitate analysis that leverages the large amount of data contained in the database. 

Because of the scale of the data stored within IBM Functional Genomics Platform, traversing across the entity tables is a non-trivial task. We used mapping tables to maintain the relationships between entities and optimize performance.  By using such mapping tables, in conjunction with DB2's columnar table store, we were able to optimize the number of joins needed to traverse the biological entities within the central dogma of molecular biology, resulting in faster queries.

\section{User Interface}
\label{sec:userInterface}
In order to allow rapid traversal across biological entities in the central dogma of molecular biology\cite{crick1970}, we have surfaced data to the user in multiple ways to accommodate differing needs. We have designed an interface and developer toolkit with two types of users in mind: computational biologists and microbiologists or biologists. Each user type has varying development skills which require different modes of access to the data. The developer toolkit to allow users to programmatically access data of this magnitude while bypassing costly biological computation. We also offer a web-based browser interface to support non-programmatic access to the data and analytics. Both of these endpoints leverage optimized Elasticsearch indices which allow for rapid retrieval and searching of all biological entities and their connected relationships.

\subsection{Developer Tools}
An important aspect of the system design is the ability to query the database for different entities based on various attributes. This database is unique due to the relationships stored between a genome accession, and its downstream gene sequences, protein sequences, and domain information (Section \ref{sec:relations}). Thus, to allow for the greatest utility of this relation feature, we have designed a method of data retrieval that allows a user to retrieve any of the entities described above (Section \ref{sec:entities}) based on its association with another entity and metadata. For example, a user can query for genes of a particular length, within a specific genus, or from a given host organism. We have allowed the user programmatic access to this data to facilitate faster queries, as well as integration with other common bioinformatic tools.

To perform analysis with IBM Functional Genomics Platform data, we have created two methods to programmatically query the database. First, we have implemented a set of generalized REST APIs that allow a user to query for specific entities based on various attributes, relationships, and metadata. These APIs are structured in a generalized manner to allow the user to retrieve the data that specifically meets their needs. For instance, a user can query for proteins within a particular genus and filter by host organism, and thus retrieve data specific to their topic without any post-processing. Additionally, the same type of queries can be produced through our Python SDK. The functions within the SDK follow the same structure as the APIs, providing consistency across all platforms. 

\begin{figure}[!ht]
\begin{center}
\includegraphics[trim=0cm 0cm 0cm 0cm, width=1\columnwidth]{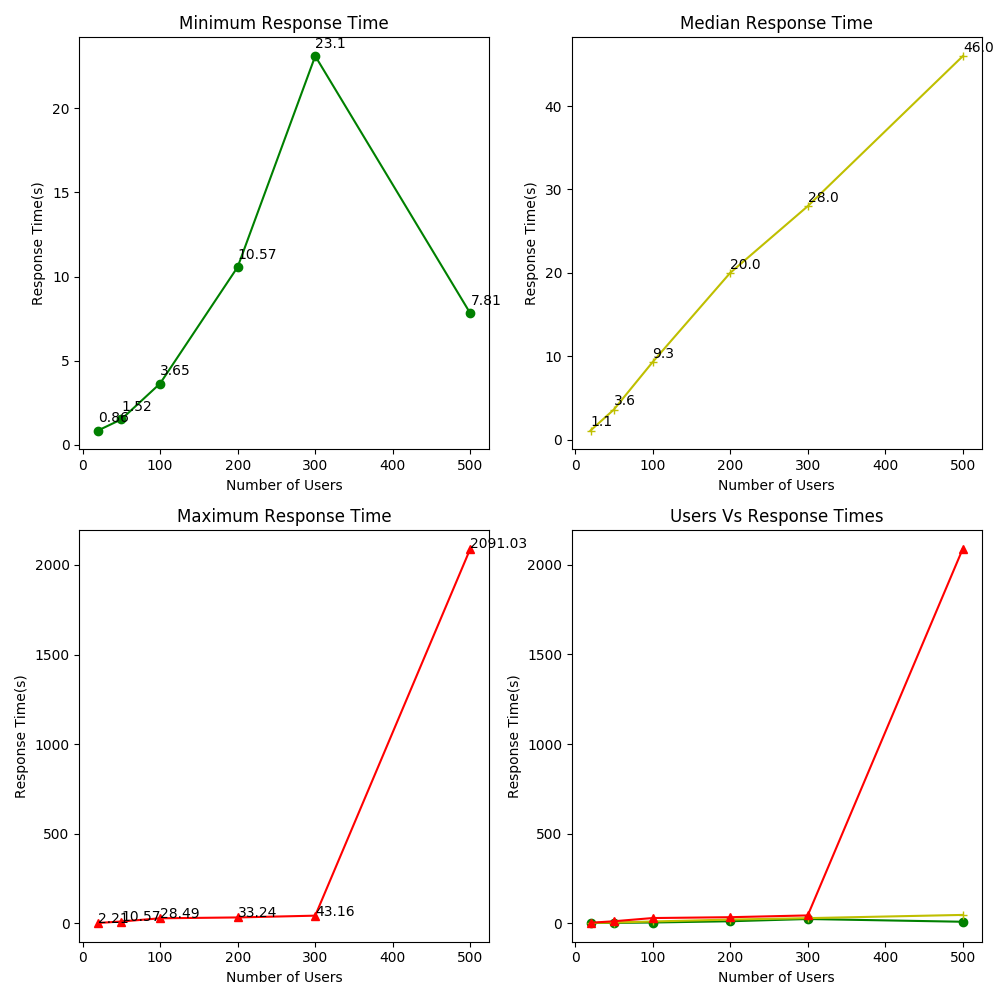}
\end{center}
\caption{Describes the speed at which results are returned for a variety of queries and how the system scales with an increase in users. This figure reports the minimum response time, average response time, maximum response time, and the comparison between the three measures given the increase in users.}
\label{fig:loadTesting} 
\end{figure}

Through the SDK and APIs, results can be returned as a JSON structure, data frame, or FASTA format in order to support a variety of downstream use cases. The JSON and Pandas dataframe formats are compatible with most machine learning algorithms. By integrating the IBM Functional Genomics Platform queries through a Python SDK, we enable a user to retrieve data within a format and framework that supports most commonly used machine learning analytics. Furthermore, many bioinformatic tools use a fasta file format as input, so this option can be used for further downstream analysis with a variety of tools. From Figure \ref{fig:loadTesting}, we highlight the speed at which the system returns results from the SDK and APIs. This shows that data is returned on average within a minute when there are 500 concurrent users. We are able to achieve this highly scalable performance due to the relational database schema (Section \ref{sec:schema}) and the usage of ElasticSearch to power search results.

We also provide two Docker containers to help users leverage the database quickly and easily in their development workflow and environment. One container has the OMXWare Python SDK package, its dependencies, and common utilities e.g. MatPlotLib and numpy. The second container provides JupyterLab and example notebooks using our SDK. Users can find complete documentation for our Docker containers, API services and SDK functions in the Develop section of the IBM Functional Genomics Platform webpage. We have also integrated BLAST within the system to allow a user to compare their own nucleotide or amino acid sequences against the IBM Functional Genomics Platform corpus. This capability is accessible through the IBM Functional Genomics Platform Hub (Section \ref{sec:OMXWareHub}), where the user inputs a sequence that gets BLASTed against pre-built BLAST databases that reflect the data. Thus, we have already performed the computationally intensive task of building the BLAST database with the data from our system, and a user can quickly identify the similarity between their sequence and the available data. In these way, we aim to make the database both programmatically usable and compatible with existing, prevalent bioinformatic pipelines.  

\subsection{IBM Functional Genomics Platform Hub}
\label{sec:OMXWareHub}
A web portal serves as the central hub for IBM Functional Genomics Platform resources. Here, a user can search for data, find summary statistics including counts of entity types within the database, read community engagement posts including ongoing research findings in the News and Learn sections, and discuss technical issues. Additionally, we have developed an Explore page that allows the user to experience the relationships captured in the database through a gallery of graphs and visualizations. 

A key aspect of the IBM Functional Genomics Platform Hub is the capability to perform a keyword search across any biological entity within the database--- over 300 million sequences. This is in contrast to many other repositories that only support targeted searching within a single biological entity. By leveraging the relational database, we are able to easily traverse all of the many-to-many connections across biological entities. Due to the optimization of the database and the use of Elasticsearch, we are able to return tens of thousands of matching results in this manner within milliseconds, along with all of their related entities. The various entities are also returned with their appropriate metadata, curated from NCBI, in order to provide necessary information for biological analysis.

\subsection{IBM Functional Genomics Platform Apps}
\label{sec:omxware_apps}

Collaborative development of applications on top of the IBM Functional Genomics Platform database is supported through both the user interface and GitHub. In the Develop section of IBM Functional Genomics Platform Hub, we showcase several applications that have been built using IBM Functional Genomics Platform data. A user can also upload an application with the option to allow others to use their tool or release source code. This is meant to foster community development and reuse of novel biological tools that are built upon the database and infrastructure. A larger-scale application platform to support computationally complex applications created by users is part of ongoing work.

\subsubsection{Example Application - Feature Discovery}
\label{sec:exampleApp}
In this section, we describe an approach for the discovery of biological features relevant to a particular phenotype such as virulence. The phenotype for virulence includes co-occurring cellular structures such as pilli, flagella, or secretions systems as well as the ability to produce toxins or transport important ionic metals. In general, functionality within a phenotype requires several distinct proteins. For this application, we leverage the fact that bacterial genomes are gene dense and that genes expressed together often exist near one other in gene-clusters. We built this application on top of IBM Functional Genomics Platform to expand the scope of feature discovery beyond an initial gene set for a more comprehensive and robust phenotype signature. Beginning with a set of input genes, proteins, or their domains, we first find all gene features with the observed domain architecture and then perform a relative position analysis to find additional neighboring genes or proteins to enhance sensitivity and specificity of the phenotype signature.

In a specific use case focused on virulence, we began with a list of 34 unique gene names (introduced by a domain expert) which the corresponded to 146 related protein domain codes and domain architectures (a combination of domain codes) within the system. This set of domain combinations became the initial feature set for analysis. We then found all proteins within the database that contain any of these domain combinations, which resulted in 7,710,132 unique protein sequences. We call this protein set "candidate proteins," as they are the proteins that are candidates for analysis after filtering steps. We then found all "candidate genomes" containing one or more of these candidate proteins. In order to prioritize phenotype enrichment (i.e. genomes with the most virulence features), we include only candidate genomes that contain more than ten of the original domains and call them the "in-group genomes." All the other candidate genomes form the "out-group genome" set. This is to avoid including genomes that may not display the virulence phenotype in the analysis. For all candidate proteins, we calculated the log-scale ratio of the number of occurrences within an in-group genome to the number of occurrences in an out-group genome. We then select only those candidate proteins that are at a log-scale ratio of 0 or higher (i.e. selecting those that are present in the in-group more frequently than the out-group) and called this set "pivot proteins." Details about the count of each entity before and after filtering can be found in Table \ref{tab:dataSelectionCount}.

\begin{table}[ht]
\centering
\caption{Counts of Entities during Data Selection}
\label{tab:dataSelectionCount}
\begin{tabular}{ll}
\multicolumn{1}{l}{Entity Description} & \multicolumn{1}{l}{Count} \\ \hline
Input Genes                 & 34        \\
Domain Codes                & 146       \\
Candidate Protein (unique name) & 4,272 \\
Candidate Protein (unique sequence) & 7,710,132 \\
Candidate Genome           & 193,223 \\
In-Group Genome             & 3,805 \\
Pivot Protein (unique name) &3,993 \\
Pivot Protein (unique sequence) & 7,691,624 \\
\end{tabular}
\end{table}

\begin{figure}[!ht]
\begin{center}
\includegraphics[trim=0cm 0cm 0cm 0cm, width=1\columnwidth]{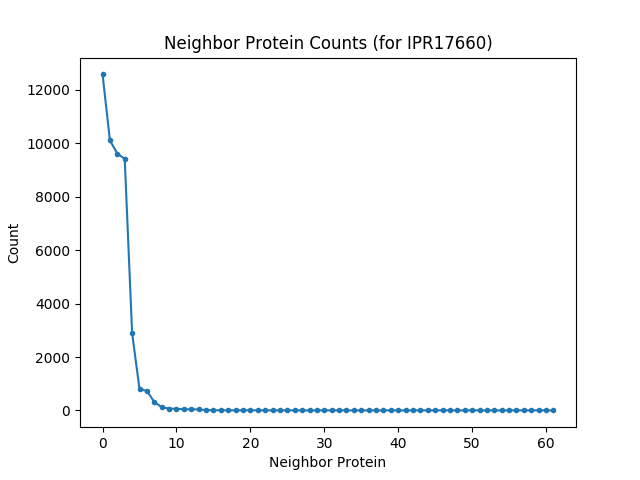}
\end{center}
\caption{Count of neighbor proteins for IPR17660. This figure shows the number of times each protein name occurs as a neighbor of IPR17660 across all genomes. }
\label{index} 
\end{figure}

The pivot proteins are then used to find adjacent (nearest neighbor) proteins on the same genome(s). We retrieve the neighbor proteins from IBM Functional Genomics Platform using the Prokka relative index $i$ of the pivot protein on a given genome contig and return the two proteins at location $i+1$ and $i-1$ on the same contig. As discussed in Section \ref{sec:curation}, the increased continuity of our genome assemblies allows for high accuracy when investigating the relative position of proteins (or genes). Using this relative index, proteins located on the end of a contig will yield some false neighbors, but for high-quality assemblies with fewer long contigs, these false neighbors are rare and easily filtered based on frequency. Evidence for index consistency is provided by Figure \ref{index}, where we plot the counts for each putative neighbor of one input domain code. We can see that 10 neighbor proteins are identified consistently across all genomes, with a long tail of proteins that are identified only once. By filtering out all neighboring proteins that occur with low frequency, we can find a consistent set of neighbor proteins using this relative index.

If the neighbor protein was also one of the original pivot proteins, we call this a co-pivot protein. If the neighbor protein was not included in the original set, this is a discovered protein. Using this discovered protein set, we can augment the known proteins that contribute to the virulence phenotype signature. Using this procedure, we observed 114 uniquely named co-pivot proteins and 66 uniquely named discovered proteins for virulence across all genomes in the database. The results are summarized in Table \ref{tab:neighborIdentificationCount}.

\begin{table}[ht]
\centering
\caption{Counts of Proteins found in Neighbor Identification}
\label{tab:neighborIdentificationCount}
\begin{tabular}{lll}
\multicolumn{1}{l}{Entity Description} &
\multicolumn{1}{l}{Name Count} &
\multicolumn{1}{l}{Sequence Count} \\ \hline
Co-Pivot Protein    & 114   & 284        \\
Discovered Proteins & 66    & 84       \\
\end{tabular}
\end{table}

The structure of the IBM Functional Genomics Platform database is crucial to this analysis as it was possible to quickly traverse from domains to proteins to genomes to identifying adjacent proteins with significant data expansion as these connections are traversed. Without a relational database to facilitate these connections, such analyses would not be possible on a large scale. The entire pipeline would have to be re-run for every input set, as this type of analysis requires the relationships between genomes down to functional domains. Furthermore, if the entities found were stored in files, rather than a relational database, the retrieval of the connections would require a large computation time simply to parse through each file. As described in Section \ref{sec: systemDetails}, we processed the original and intermediate files using a combination of bare metal and virtual machines, ultimately generating over 160TB of data. Therefore, it is clear that this type of neighbor protein analysis is not computationally tractable for a single case without the use of a pre-computed relational database. 

Furthermore, the application has been contributed back to the community and deployed as a generalized micro-service to allow users to input any set of domains or proteins relevant to their study. The structure of the APIs created for this example application support feature discovery for any phenotype of interest. We have pre-computed example inputs for several phenotypes, including virulence, to allow for exploration of this application. The application can be accessed through the IBM Functional Genomics Platform Develop page and can be launched from the user interface for ease of use.

\section{Conclusion}
\label{sec:conclusion}
A relational database linking genotype to phenotype across over 1,000 genera of microbial life transforms the way in which bioinformatic questions can be asked and answered today. The database grows only at the rate new features are discovered in real-time as genome assemblies, annotations, and relations are continually updated and stored uniquely in a compute cloud. Previously observed biological entities and their relations need not be recomputed when identified again as new data is added to the system.  This property is essential to keeping pace with the rate of newly sequenced samples in the age of big data. Most importantly, the computational work required to build the database itself means the answers to many biological questions have been pre-computed and can be retrieved by simply querying the database. The study and discovery of co-occurring virulence features described above is an example of this property. For this research, there was no need to build static indices or workflows as required for traditional bioinformatic tools e.g. BLAST\cite{kent2002}, Kraken\cite{wood2014}, or Bowtie2\cite{Bowtie:2018}, but instead, a new query simply had to be executed to map targeted proteins and then to find their neighbors. The capability has now been expanded to a general use micro-service for phenotype signature discovery to further leverage the pre-computed relationships and genome structure. This significant paradigm shift supports streaming of biological data as biological entities are updated in real-time instead of the traditional method of building references from static files.

The database and system reported here demonstrate how bioinformatics can scale in the age of large sequence data. At the time of this report, the database contains extensive metadata related to over 200,000 high-quality bacterial genomes, over 68 million unique genes, over 52 million unique proteins, and over 239 million unique protein domains. Using the pipeline described above, we annotate each genome to identify its gene sequences, protein sequences, and ultimately the functional domains, and maintain each unique entity along with connections to its parent genome. The relationships between genomes, genes, proteins, and protein domains are central to connecting genotype to phenotype. The IBM Functional Genomics Platform API services and SDK allow one to build and test new applications for diverse research interests. 

\subsection{Ongoing work}

In order to provide ease of use as a research platform, we are taking key steps to augment the IBM Functional Genomics Platform system. Currently, we are in a limited release and plan to expand user access and capabilities in the coming year. We are developing an application framework to allow users to contribute applications to the system and to leverage a broad user community. As the application framework expands, our goal is to increase the benefit to the scientific community by allowing an individual user to make their applications accessible through the platform. Additionally, in order to further support microbial analysis, we are developing capabilities to assist users in uploading and analyzing their own sequence data. It is important to allow researchers to compare their microbial data to the IBM Functional Genomics Platform reference data. To this end, we are developing a method for data upload into new schema instances and to allow for interoperability between user-generated data and IBM Functional Genomics Platform reference data.

\ifCLASSOPTIONcompsoc
  \section*{Acknowledgment}
\else
  \section*{Acknowledgment}
\fi

The authors would like to acknowledge Dr. C. A. Elkins and his team at
US Centers for Disease Control and Prevention for inspiring the prototype Virulence Feature Discovery App and for providing the first features for testing. We would also like to acknowledge Dr. N. Haiminen and Dr. L. Parida of IBM Research for helpful discussion and insights into new applications of IBM Functional Genomics Platform.

\ifCLASSOPTIONcaptionsoff
  \newpage
\fi



\bibliographystyle{IEEEtran}
\bibliography{IEEEabrv,paper_refs}
%



%

\newpage 
\begin{IEEEbiographynophoto}{Edward E. Seabolt}
An experienced Software Engineer with a demonstrated history of architecting,
developing, and leading large and complex software systems. Highly skilled in the areas of Cloud, BigData, Database and distributed system design. Currently the Chief Architect and Lead Developer for IBM Functional Genomics Platform, a cloud data platform for the study of microbial life at scale containing over 300 million genomes, genes, proteins and domains. Ed holds a Masters in Chemistry from Oklahoma State University and served in the United States Marine Corps Reserve.
\end{IEEEbiographynophoto}

\begin{IEEEbiographynophoto}{Gowri Nayar}
Gowri Nayar is a research software engineer at IBM Research, Almaden for the IBM Functional Genomics Platform, focusing on developing machine learning analytics for genomic datasets, identifying key characteristics of genes and proteins. Her interest is in using foundational concepts of machine learning to develop methods for interdisciplanary problems. She has a Masters Degree from Georgia Institute of Technology in Computer Science. 
\end{IEEEbiographynophoto}

\begin{IEEEbiographynophoto}{Harsha Krishnareddy}
Harsha Krishnareddy is a Software Engineer for the IBM Functional Genomics Platform . His interest is in the emerging field of Knowledge Graphs, AI and Natural Language Processing to build intelligence software systems. He has a Masters Degree in Information Systems and Management from Illinois Institute of Technology (IIT) Chicago.
\end{IEEEbiographynophoto}

\begin{IEEEbiographynophoto}{Akshay Agarwal}
Akshay is a Software Designer in the AI and Heathcare Life Sciences group at IBM Research Almaden. He graduated with a Masters in CS from Georgia Tech and since have been working in IBM Research, where he works on IBM Functional Genomics Platform, the largest database on bacterial life. His research involves studying antimicrobial resistance and how to predict and circumvent it.
\end{IEEEbiographynophoto}

\begin{IEEEbiographynophoto}{Kristen L. Beck}
Kristen L. Beck, PhD is the Lead Bioinformatician and Research Staff Member in the AI and Cognitive Solutions team at IBM Research - Almaden. Her work involves developing multi-omics approaches for human health and the life sciences including the study of anomaly detection in food microbiomes, multimodal fusion of oncology data, and microbial pathogenicity.
\end{IEEEbiographynophoto}

\begin{IEEEbiographynophoto}{Igancio Terizzano}
Ignacio Terrizzano is a Sr. Software Research Engineer with over 30 years of industry experience.  Currently with IBM Research in San Jose,  he has led, designed, and developed several systems in varied disciplines such as bioinformatics, artificial intelligence, services science, and business architecture.  Ignacio has rich experience and deep interest in complex modeling, and holds several patents in this area.  Ignacio holds a Master’s Degree in Computer Science from the University of Colorado, and a Bachelor’s Degree in Computer Science from the University of Santa Clara.
\end{IEEEbiographynophoto}

\begin{IEEEbiographynophoto}{Eser Kandogan}
Eser Kandogan is currently a staff software engineer at Megagon Labs. Prior to Megagon he was a research staff member at IBM Almaden Research Center and conducted research on visual analytics, human-computer interaction,  information visualization and search. He holds a Ph.D. degree from University of Maryland in Computer Science.
\end{IEEEbiographynophoto}

\begin{IEEEbiographynophoto}{Mark Kunitomi}
Mark Kunitomi is a Research Staff Member at IBM Research. He has expertise in Genomics, Bioinformatics, and Microbiology. Mark earned his Ph.D. at The University of California, San Francisco in Biochemistry and Molecular Biology.
\end{IEEEbiographynophoto}

\begin{IEEEbiographynophoto}{Mary Roth}
Mary Roth is the VP of Engineering Operations at Couchbase. She was previously a senior manager and principal researcher IBM, earning the title of Distinguished Engineer, reserved for the top 1\% of IBM's 200,000+ technical employee population.  Mary is an industry recognized, award-winning technical leader with expertise in database systems, data integration, data governance, and analytics. She has a proven track record of results-driven innovation and team leadership, including 3 major concept-to-product launches, 20+ patents and 25+ academic publications. 
\end{IEEEbiographynophoto}

\begin{IEEEbiographynophoto}{Vandana Mukherjee}
Vandana Mukherjee, Ph.D, a senior research manager at IBM Almaden in Health Care Life Sciences with teams focused on AI, Natural Language Processing, Medical Imaging, Health Content Extraction and Genomics. She has been in IBM for over 20 years and also has experience in IBM systems and microelectronics.
\end{IEEEbiographynophoto}

\begin{IEEEbiographynophoto}{James H. Kaufman}
James H. Kaufman is a scientist in the AI and Cognitive Software organisation at the IBM Almaden Research Center in San Jose, CA. He received his B.A. in Physics from Cornell University and his Ph.D. in Physics from the  University of California, Santa Barbara. He is a Fellow of the American Physical Society, a Distinguished Scientist of the Association for Computing Machinery, and an IBM Distinguished Research Staff Member. He is also Eclipse project co-lead for the open source SpatioTemporal Modeler. He is currently serving as chief scientist a project using molecular data to improve our understanding of host-microbe interactions and disease phenotype.
\end{IEEEbiographynophoto}







\end{document}